\documentclass{article}
\usepackage{ijcai25}

\usepackage{times}
\usepackage{soul}
\usepackage{url}
\usepackage[hidelinks]{hyperref}
\usepackage[utf8]{inputenc}
\usepackage[small]{caption}
\usepackage{graphicx}
\usepackage{amsmath}
\usepackage{amsthm}
\usepackage{booktabs}
\usepackage{algorithm}
\usepackage{algorithmic}
\usepackage[switch]{lineno}

\usepackage{algorithm}
\usepackage{algorithmic}
\usepackage{amsmath}
\usepackage{amsthm}
\usepackage{float} %设置图片浮动位置的宏包
\usepackage{centernot}
\usepackage{multirow} %设置表格
\usepackage{array}
\usepackage{subfigure}
\usepackage{xcolor}
\usepackage{color}
\usepackage{float}
\usepackage{amsfonts,amssymb}
\usepackage{subfigure}

\newtheorem{definition}{Definition}

\title{Interaction-Data-guided Conditional Instrumental Variables for Debiasing Recommender Systems}

\author{
Zhirong Huang$^{1,2}$\and
Debo Cheng\textsuperscript{*}$^{3,4}$\and
Lin Liu$^{4}$\and
Jiuyong Li$^{4}$\and
Guangquan Lu$^{1,2}$\And
Shichao Zhang\textsuperscript{*}$^{1,2}$
\affiliations
$^1$Key Lab of Education Blockchain and Intelligent Technology, Ministry of Education, Guangxi Normal University, Guilin, 541004, China\\
$^2$Guangxi Key Lab of Multi-Source Information Mining and Security, Guangxi Normal University, Guilin, 541004, China\\
$^3$School of Computer Science and Technology, Hainan University, Haikou, Hainan, 570228, China\\
$^4$UniSA STEM, University of South Australia, Mawson Lakes, Adelaide, Australia\\
\emails
huangzr@stu.gxnu.edu.cn, chengdb2016@gmail.com, \\ \{zhangsc,lugq\}@mailbox.gxnu.edu.cn, \{Lin.Liu, Jiuyong.Li\}@unisa.edu.au
}

\begin{document}

\maketitle

\begin{abstract}
    It is often challenging to identify a valid instrumental variable (IV), although the IV methods have been regarded as effective tools of addressing the confounding bias introduced by latent variables. To deal with this issue, an Interaction-Data-guided Conditional IV (IDCIV) debiasing method is proposed for Recommender Systems, called IDCIV-RS. The IDCIV-RS automatically generates the representations of valid CIVs and their corresponding conditioning sets directly from interaction data, significantly reducing the complexity of IV selection while effectively mitigating the confounding bias caused by latent variables in recommender systems. Specifically, the IDCIV-RS leverages a variational autoencoder (VAE) to learn both the CIV representations and their conditioning sets from interaction data, followed by the application of least squares to derive causal representations for click prediction. Extensive experiments on two real-world datasets, Movielens-10M and Douban-Movie, demonstrate that IDCIV-RS successfully learns the representations of valid CIVs, effectively reduces bias, and consequently improves recommendation accuracy.
    % The source code for IDCIV-RS is available at: \url{https://anonymous.4open.science/r/IDCIV-RS}.
\end{abstract}
\footnotetext{$^*$Corresponding author}
\section{Introduction}
With the rapid development of the Internet, the amount of information has exploded, making it increasingly difficult for users to sift through vast amounts of data to find content that aligns with their preferences~\cite{luo2025fairgp,gao2021advances,zhang2021challenges}. Recommender systems have emerged as a critical solution to this problem by analysing user behaviour data to deliver personalised recommendations, thereby enhancing user engagement and satisfaction~\cite{wang2020causal}. Recommender systems have become an integral part of many digital platforms, finding extensive applications across various domains, including e-commerce \cite{shoja2019customer}, streaming media \cite{gomez2015netflix}, and social networks \cite{liao2022sociallgn}, significantly improving information retrieval efficiency and user experience.

The performance of recommendation systems is often affected by latent confounders that are neither directly observable nor reflected in historical user-item interactions. For instance, factors such as social influence or peer preferences can significantly shape user behavior but are rarely captured in logged data. This misalignment may cause to recommendation systems misinterpret user intent, ultimately degrading recommendation quality. Effectively addressing latent confounders is thus vital for enhancing both the accuracy and robustness of recommender systems.

% However, the performance of existing recommendation systems is often hindered by latent confounders that are not directly observable or accounted for in the historical user-item interaction data. For example, a user's social network or peer influence can shape their preferences for certain items, yet such interactions may not be fully captured by user behaviour data. As a result, recommendation algorithms may misinterpret users' true preferences, leading to a decline in recommendation quality. Addressing the impact of latent confounders is therefore crucial for improving the accuracy and effectiveness of recommender systems.

\begin{figure}[t]
\begin{center}
    \subfigure[]{
        \includegraphics[width=0.245\columnwidth]{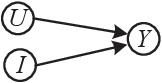}
        }
        \subfigure[]{
        \includegraphics[width=0.245\columnwidth]{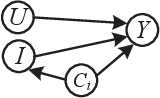}
        }
        \subfigure[]{
        \includegraphics[width=0.30\columnwidth]{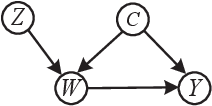}
        }
    \caption{Three causal graphs illustrate the assumptions underlying existing work. The variables are defined as follows: $U$: user preferences, $I$: displayed items, $Y$: user feedback or outcome variables, $C_i$: confounding factors affecting items, $C$: confounding factors, $Z$: instrumental variables, $W$: treatment variables (i.e., embeddings of user-item pairs). (a) The causal graph represents the traditional recommendation model; (b) the causal graph represents the debiasing method using causality; (c) the causal graph representation of the debiasing method using standard IV.}
\label{Fig1} 
\end{center}
\end{figure}

User behaviour data is crucial for recommender systems in predicting user preferences. Existing models often assume this data is unbiased and accurately reflects user preferences, meaning the data accurately reflects user preferences~\cite{lan2024contrastive}. Based on this assumption, many methods have been proposed, such as Matrix Factorisation (MF)~\cite{koren2009matrix} and Neural Network-based Collaborative Filtering (NCF)~\cite{he2017neural}. These methods achieve the goal of predicting user preferences by fitting user behaviour data, as depicted in the causal Directed Acyclic Graph (DAG)~\cite{pearl2009causality,cheng2022ancestral} shown in Figure~\ref{Fig1} (a), which illustrates the causal relationships between user preferences $\left(U\right)$ and user feedback $\left(Y\right)$, as well as between displayed items $\left(I\right)$ and $Y$. However, in the real-world, user behaviour is inevitably influenced by various unobservable confounding factors, such as item popularity (e.g., frequent recommendation of an item limits users' choices) and user psychology (e.g., choosing to watch a movie for socialising)~\cite{chen2023bias}. These factors introduce biases like popularity and conformity bias, leading models to learn false correlations and reducing their ability to accurately predict preferences.

% Causal inference techniques have been utilised to mitigate bias in recommender systems~\cite{wang2020causal,gao2024causal}. These methods leverage domain knowledge to design causal DAGs that represent the data generation process, identify specific biases, and develop models to address them. For example, Zhang et al.~\cite{zhang2021causal} proposed a novel causal graph (Figure~\ref{Fig1} (b)) that analyses how item popularity affects recommendations. Based on this graph, they introduced the Popularity-bias Deconfounding and Adjusting (PDA) training paradigm to mitigate popularity bias. Similarly, Zheng et al.~\cite{zheng2021disentangling} examined the impact of users' conformity mentality and proposed the Disentangling Interest and Conformity with Causal Embedding (DICE) method to address conformity bias. While these causal graph-based approaches have shown some success, they rely on the assumption that the data generation process strictly follows the proposed causal graph. If real-world conditions deviate significantly from these assumptions, their effectiveness may be limited~\cite{cai2024mitigating}. Moreover, the presence of numerous confounding factors in real-world scenarios makes it challenging to satisfy these assumptions.
Causal inference~\cite{zhang2024learning,li2024contrastive} has been applied to reduce bias in recommender systems by designing causal DAGs to model data generation, identify biases, and guide model design~\cite{wang2020causal,gao2024causal}. For instance, Zhang et al.~\cite{zhang2021causal} introduced a causal graph to analyse item popularity's impact and proposed the PDA training paradigm to correct popularity bias. Similarly, Zheng et al.~\cite{zheng2021disentangling} addressed conformity bias with the DICE model, disentangling user interest from conformity. However, these methods rely on the assumption that the real data follows the designed causal graphs, which may not hold in practice~\cite{cai2024mitigating}. Complex real-world confounding makes such assumptions difficult to satisfy, potentially limiting the models' effectiveness.

% Latent confounders pose a significant challenge in debiasing recommender systems. These latent confounders (a.k.a. unobserved or unmeasured variables) affect both the treatment (e.g., the recommendation process) and the outcome, leading to biased recommendations. Instrumental Variables (IVs) are often used to eliminate bias caused by latent variables~\cite{caner2004instrumental,pearl2009causality}, e.g., $Z$ is an IV as shown in Figure~\ref{Fig1} (c). A standard IV must meet three conditions~\cite{pearl2009causality}: (i) relevance to the treatment variable; (ii) an exclusive impact on the outcome through the treatment; and (iii) no shared confounders with the outcome. Several debiasing methods for recommender systems based on standard IVs have been developed to address latent confounding factors without relying on strict causal graph assumptions~\cite{si2022model,si2023enhancing,si2023search}. For instance, IV4Rec~\cite{si2022model} uses self-collected search data as an IV to effectively mitigate latent confounding. However, verifying the last two conditions of a standard IV solely through data is infeasible~\cite{brito2012generalized,cheng2023causal}, making the identification of valid IVs a significant challenge.
Latent confounders, which are unobserved variables that simultaneously affect both the treatment (e.g., recommendation process) and the outcomes, present significant challenges for debiasing recommender systems. Instrumental Variables (IVs) are a common solution to this issue~\cite{caner2004instrumental,pearl2009causality}, with a valid IV (e.g., $Z$ in Figure~\ref{Fig1}(c)) satisfying three key criteria~\cite{pearl2009causality}: (i) relevance to the treatment variable; (ii) an exclusive impact on the outcome through the treatment; and (iii) no shared confounders with the outcome. Several IV-based methods have been proposed for recommendation settings to reduce latent bias without relying on strict causal graph assumptions~\cite{si2022model,si2023enhancing,si2023search}. For example, IV4Rec~\cite{si2022model} leverages self-collected search data as an IV to effectively address latent confounding. However, verifying the latter two IV conditions from observational data alone is practically infeasible~\cite{brito2012generalized,cheng2023causal}, making the identification of valid IVs particularly challenging.

\begin{figure}[t]
    \centering
    \includegraphics[width=0.5\columnwidth]{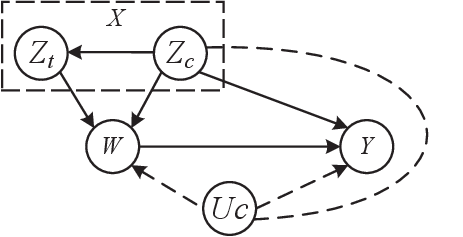}
    \caption{A causal DAG illustrating our proposed IDCIV-RS method for discovering CIVs and their conditional sets from observational data. $W$, $Y$, $X$, and $U_c$ are the treatment, outcome, set of measured pretreatment variables, and latent confounders between $W$ and $Y$, respectively. $Z_t$ and $Z_c$ denote the representations of the CIVs and their conditional sets learned from $X$.}
    \label{Fig2}
\end{figure}

In causal inference, researchers have used conditional IVs (CIVs) to tackle the limitations of standard IVs~\cite{pearl2009causality,brito2012generalized,cheng2024instrumental,cheng2023learning}. CIVs offer more relaxed application conditions than standard IVs. Recently, Cheng et al.~\cite{cheng2023causal} developed a CIV method (CIV.VAE) based on the variational autoencoder (VAE)~\cite{kingma2013auto,scholkopf2021toward,scholkopf2022causality} model, which generates CIVs and their conditional sets from data, significantly relaxing the constraints of standard IVs. However, this method is designed for tabular data, and no attempt has been made to adapt it to interactive data in recommender systems. 

To address the challenge of using CIV in interactive data, we first propose a causal DAG, as shown in Figure~\ref{Fig2}, to represent the causal relationships between observed and latent variables in interaction data within recommender systems. Building on this causal DAG, we develop an interaction-data-guided conditional IV (IDCIV) debiasing method for recommender systems, called IDCIV-RS. Specifically, IDCIV-RS uses the embeddings of user-item pairs (including users with the selected items, i.e., positive samples, and users with their pre-selected items, i.e., negative samples) as the treatment variable $W$, the user feedback as the outcome $Y$, and the user interaction data (only positive samples) as the pretreatment variable $X$. We assume that at least one CIV exists within $X$, capturing latent information such as user search behaviours that lead to interactions. The assumption is reasonable because in real-world scenarios, users are often influenced by certain external factors that cause their interactions to be not fully consistent with their true preferences. For example, users may interact with items because of certain external incentives (e.g., search recommendations, promotions, etc.) even though these items do not exactly match their true preferences. IDCIV-RS employs a VAE to generate the representations of the CIV and its conditional set from $X$, denoted as $Z_t$ and $Z_c$, as shown in Figure~\ref{Fig2}. The contribution of our work is summarised below:
\begin{itemize}
    \item We propose a novel causal DAG to represent the causal relationships between observed and latent variables in interaction data within recommender systems. 
    \item We develop an interaction-data-guided conditional IV (IDCIV) debiasing method for recommender systems, called IDCIV-RS, for learning representations of CIVs and its conditional sets under the proposed causal DAG. To the best of our knowledge, this is the first work to generate representations of CIVs and their conditional sets from interaction data for mitigating bias in recommender systems.
    \item Extensive experiments on two real-world datasets demonstrate that IDCIV-RS achieves optimal debiasing results and recommendation performance compared to state-of-the-art causal debiasing methods.
\end{itemize}

\section{Related Work}
In this section, we review the recommender methods most closely related to our IDCIV-RS, including traditional recommender methods and causal recommender methods.

\subsection{Traditional Recommender Methods}
Traditional recommender methods, primarily based on Collaborative Filtering (CF), typically assume that user behaviour data is unbiased~\cite{koren2021advances}. The mainstream approach is model-based CF, which trains a model on user behaviour data to recommend items that align with user preferences. A classic method is MF~\cite{koren2009matrix}, which decomposes the user-item rating matrix to predict preferences. However, MF assumes that unselected items are incompatible with user preferences, ignoring cases where users may not have encountered those items. To address this, Rendle et al.~\cite{rendle2012bpr} proposed Bayesian Personalized Ranking (BPR), which assumes users prefer selected items over unselected ones, enhancing preference inference in MF. 
% Traditional recommender systems, mainly based on Collaborative Filtering (CF), often assume unbiased user behavior data~\cite{koren2021advances}. Model-based CF methods, such as Matrix Factorization (MF)\cite{koren2009matrix}, learn user preferences by decomposing the user-item rating matrix. However, MF treats unselected items as negative feedback, overlooking cases where users have simply not encountered those items. To overcome this, Rendle et al.\cite{rendle2012bpr} introduced Bayesian Personalized Ranking (BPR), which models user preference by assuming selected items are favored over unselected ones, thereby improving preference inference.

With the rise of deep learning~\cite{luo2024fugnn,zhang2022hyper}, He et al.~\cite{he2017neural} proposed NCF, which uses multi-layer perceptrons (MLPs) to model non-linear user preferences. To capture richer behavioural signals, Wang et al.~\cite{wang2019neural} introduced Neural Graph Collaborative Filtering (NGCF), leveraging Graph Convolutional Networks (GCNs) to embed user-item interactions. He et al.~\cite{he2020lightgcn} further simplified this with LightGCN, improving both efficiency and accuracy. However, these methods often overlook popularity bias, which can be amplified during training and skew recommendations toward popular items.
% With the rise of deep learning, He et al.\cite{he2017neural} introduced NCF, using multi-layer perceptrons (MLPs) to better model non-linear user preferences and improve recommendation accuracy. Researchers also integrated graph structures to capture richer behaviour data. Wang et al.\cite{wang2019neural} developed Neural Graph Collaborative Filtering (NGCF), leveraging Graph Convolutional Networks (GCNs) to enhance recommendations by embedding user-item interactions. Building on NGCF, He et al.~\cite{he2020lightgcn} proposed LightGCN, a simplified model that improves efficiency and performance. Despite their success, traditional methods often overlook biases (e.g., popularity bias), which can be amplified during training, leading models to overemphasize popular items.

\subsection{Causal Recommender Methods}
To mitigate biases in recommender systems, researchers have increasingly adopted causal inference techniques. Early approaches used the Inverse Propensity Score (IPS)~\cite{wang2021non,schnabel2016recommendations,bottou2013counterfactual} to reduce bias by assigning an inverse propensity score (e.g., the inverse of item popularity) to user-item interactions during training, balancing the influence of popular and less popular items. However, IPS methods often suffer from high variance and instability. Building on IPS success, researchers have explored causal graph-based methods~\cite{zheng2021disentangling,he2023addressing,zhang2021causal} that model the generation mechanisms of user behaviour. These methods design specific models to address biases like popularity and conformity bias. Yet, the presence of unobserved confounders in real-world data limits the effectiveness of these causal graph assumptions~\cite{cai2024mitigating}.

IVs are commonly used in causal inference to address confounding. Recently, methods leveraging user search data as IVs have emerged, with the IV4Rec framework by Si et al.~\cite{si2022model} being a notable example. IV4Rec uses user search data to decompose user-item representations into causal and non-causal components, addressing some limitations of causal graph-based methods and reducing bias. However, identifying valid user search data as IVs remains challenging. Unlike these approaches, our work focuses on learning the representations of CIVs and their conditional sets, which are less restrictive than standard IVs.

\section{The Proposed IDCIV-RS Method}
\begin{figure*}[h]
    \centering
    \includegraphics[width=0.8\textwidth]{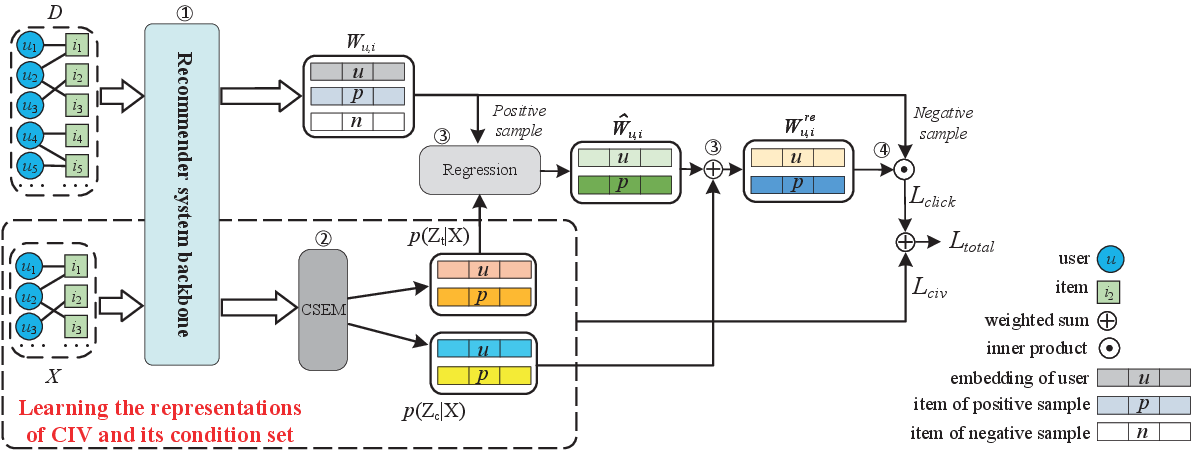}
    \caption{The overall structure of IDCIV-RS consists of four main steps, labelled 1, 2, 3, and 4 in the diagram. First, during feature encoding, IDCIV-RS uses a backbone (MF or LightGCN) to encode the input data into a latent space representation. Second, the CSEM module is used to learn the CIV $Z_t$ and its condition set $Z_c$ from the user-item interaction data $X$. Third, the representation of CIV is used to decompose the treatment variable $W_{u, i}$, obtain the causal relationship representation $\widehat{W}_{u, i}$, and fuse it with the conditional set representation $Z_c$ to get the reconstructed treatment variable $W_{u,i}^{re}$. Finally, $W_{u, i}^{re}$ is used for the click prediction.}
    \label{CIV4Rec}
\end{figure*}

In this section, we first introduce the problem definition, then explain the feasibility and rationality of our method by the causal graph, and then introduce the four main steps of our method. The overall workflow of our proposed IDCIV-RS is visualised in Figure~\ref{CIV4Rec}.
 
\subsection{Problem Definition}

In the recommender system, user behaviour data $D$ usually consists of a user set $U$ and an item set $I$. $D$ contains two parts, namely: user $u$ and selected items $p$ to form positive sample pairs, and user $u$ and pre-selected items $n$ to form negative sample pairs. The user interaction data $X$ consists of positive sample pairs derived from $D$. $X$ implicitly contains a wealth of information, including user interactions stemming from search behaviours.

In recommendation models, users and items are usually represented as low-dimensional embedding representations $W$, and the corresponding user-item pairs can be represented as $W=\{\left(w_u, w_i\right)|u\in U, i\in I\}$. However, in addition to reflecting user preferences, user behaviour data $D$ contains spurious correlations caused by various latent confounding factors $U_c$ (e.g., item exposure, conformity influence). Although existing methods have mitigated the impact of these confounding factors to some extent through causal inference techniques, they often come with strong assumptions. We aim to address this challenging problem in our work, and our problem definition is described as follows.

\begin{definition}
In a recommender system, the latent variables \( U_c \) affect the choice made and introduce bias. We assume that at least one CIV exists within $X$, capturing latent information that leads to interactions. The causal relationships between measured and latent variables are shown in Figure~\ref{Fig2}. Our goal is to learn the representations of the CIV $Z_t$ and its conditional set $Z_c$ from the user interaction data \( X \) to address the confounding biases introduced by \( U_c \). 
\end{definition}

\subsection{The Proposed Causal DAG}
In this work, we proposed a causal DAG $\mathcal{G}$ as shown in Figure~\ref{Fig2} to represent the causal relationships between the measured and latent variables. Let $\mathcal{G}_{\underline{W}}$ be the manipulated graph, obtained by deleting all arrows emerging from nodes in $W$ within $\mathcal{G}$. In $\mathcal{G}_{\underline{W}}$, $Z_t$ and $W$ are d-connected when conditioned on $Z_c$ because of the existence of the edge $Z_t \rightarrow W$. However, $Z_t$ and $Y$ are d-separated by $Z_c$ since $Z_t$, $Z_c$ and $U_c$ form a collider at $W$, and $Z_c$ blocks the path $Z_t\leftarrow Z_c \rightarrow Y$. Furthermore, the effect of $Z_t$ on $Y$ is mediated solely by $W$ through the causal path $Z_t\rightarrow W\rightarrow Y$. Therefore, $Z_t$ is a CIV, and $Z_c$ is its corresponding conditional set. Note that $Z_c$ may contain information about latent factors $U_c$ due to the complex relationships in interactive data, as indicated by the dashed line between $U_c$ and $Z_c$~\cite{wu2022instrumental,cheng2024conditional}. Many existing works have shown that confounding factors can affect recommendation outcomes, specifically $Z_c \rightarrow Y$~\cite{si2022model,si2023enhancing,si2023search}. Furthermore, since $Z_t$ and $Z_c$ are derived from the user interaction data $X$, they will affect $W$, i.e., $Z_t\rightarrow W$ and $Z_c\rightarrow W$ in $\mathcal{G}$. 

Based on the proposed causal DAG, we present IDCIV-RS, an interaction-data-guided CIV debiasing method for recommender systems. IDCIV-RS offers two key advantages over existing approaches: it eliminates the need for domain-specific IV specification and leverages CIVs, which provide a more general framework for mitigating confounding bias from latent factors.
% Based on the proposed causal DAG, we introduce an interaction-data-guided CIV debiasing method for recommender systems, called IDCIV-RS. The advantages of IDCIV-RS over existing debiasing methods are twofold: it does not require the specification of IVs through domain knowledge, and it uses CIV, which is more general than IV-based methods for addressing confounding bias caused by latent factors.

\subsection{The Concepts of Treatment Variable, CIV and Its Conditional Set}
We define the concepts for the treatment variable, the CIV, and its conditional set based on user behaviour data. The treatment variable $W_{u, i}$ using user-item pairs is defined as:
\begin{equation}
    W_{u,i} = \{\left(w_u, w_i\right)|u \in U, i \in I\},
\end{equation}
where $w_u$ and $w_i$ represent the embeddings of user $u$ and item $i$, respectively. The concepts for the representations of CIV $Z_t$ and its conditional set $Z_c$ from the user interaction data $X$ are as follows:
\begin{equation}
    Z_t = \{\left(z_{t_u},z_{t_p}\right)|u, p \in X\},
\end{equation}

\begin{equation}    
    Z_c = \{\left(z_{c_u},z_{c_p}\right)|u, p \in X\},
\end{equation}
where $z_{t_u}$ and $z_{c_u}$ indicate the representations of the CIV of the user $u$ and its conditional set, respectively, $z_{t_p}$ and $z_{c_p}$ denote the representations of CIV of the corresponding item $p$ and its conditional set, respectively. To estimate the causal effects of users and items on $Y$ (user feedback), it is necessary to construct the CIV and its conditional set for the user $u$ and item $p$, respectively.

% \subsection{Get CIV and its Condition Set}
\subsection{Learning the Representations of CIV and Its Condition Set}
\begin{figure}[htbp]
    \centering
    \includegraphics[width=0.9\columnwidth]{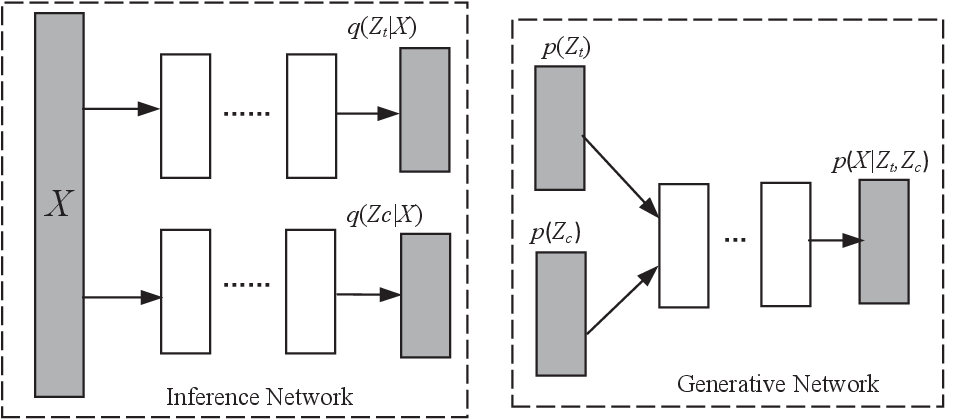}
    \caption{The CSEM components used to learn the latent representation of CIV $Z_t$ and its conditional set $Z_c$ consist of an inference network and a generation network. The grey boxes represent samples drawn from the corresponding distribution, and the white boxes represent the neural network.}
    \label{vae}
\end{figure}
In our IDCIV-RS framework, we employ a VAE structure as the generative model to generate the representations of the CIV $Z_t$ and its conditional set $Z_c$~\cite{kingma2013auto,sohn2015learning} from the user interaction data $X$, referred to as the CIV and conditional Set Extraction Module (CSEM), as shown in Figure~\ref{vae}. We use the inference and generation networks of VAE to approximate the posterior distributions $p\left(Z_t|X\right)$ and $p\left(Z_c|X\right)$ for the two representations $Z_t$ and $Z_c$.

In the inference network, we use two independent encoders to learn the posterior distributions $q\left(Z_t|X\right)$ and $q\left(Z_c|X\right)$. The variational approximations of their posterior distributions are as follows:
\begin{equation}
    q\left(Z_t|X\right) = \prod^{D_{Z_t}}_{m=1}\mathcal{N}\left(\mu = \widehat{\mu}_{Z_{t_m}},\sigma^2 = \widehat{\sigma}_{Z_{t_m}}^2 \right),
\end{equation}
\begin{equation}
    q\left(Z_c|X\right) = \prod^{D_{Z_c}}_{m=1}\mathcal{N}\left(\mu = \widehat{\mu}_{Z_{c_m}},\sigma^2 = \widehat{\sigma}_{Z_{c_m}}^2 \right),
\end{equation}
where $\mu$ and $\sigma$ are the mean and variance of the Gaussian distribution captured by neural networks. It is worth noting that $Z_t$ and $Z_c$ are composed of pairs of user and item samples, so each $q\left(Z_t|X\right)$ and $q\left(Z_c|X\right)$ has two components: item and user.
In the generative network, the prior distribution $p\left(Z_t\right)$ follows a Gaussian distribution:
\begin{equation}
    p\left(Z_t\right) = \prod^{D_{Z_t}}_{m=1}\mathcal{N}\left(Z_{t_m}|0,1\right).
\end{equation}

The prior distribution $p\left(Z_c\right)$  is obtained based on the Conditional VAE (CVAE)~\cite{sohn2015learning} model. We use Monte Carlo (MC) sampling to condition $p(Z_c)$ on $X$:
\begin{equation}
    Z_c \sim p(Z_c|X).
\end{equation}

Then, the decoder for $X$ is described as follows:
\begin{equation}
    p(X|Z_t,Z_c) = \prod^{D_{X}}_{m=1} p\left(X_m|Z_t,Z_c\right).
\end{equation}

%Using CVAE to condition $p(Z_c)$ on $p(Z_t)$ is a critical step because $Z_c$ is a conditional set of $Z_t$, which ensures that $Z_t$ can contain valid CIV information. 
For inference, we optimise the parameters by maximising the evidence lower bound (ELBO):
\begin{equation}
\begin{split}
    L_{civ} = & \mathbb{E}_q[\log p\left(X|Z_t,Z_c\right)] - D_{KL}[q\left(Z_t|X\right)|| p\left(Z_t\right)]\\& - D_{KL}[q\left(Z_c|X\right)||p\left(Z_c|X\right)].
\end{split}
\end{equation}

Note that using CVAE to condition $p(Z_c)$ on $X$ is a critical step for learning the representations of $Z_c$ and $Z_t$ because $Z_c$ and $Z_t$ are independent given $X$, which ensures that $Z_t$ captures the CIV information, while $Z_c$ capture the confounding information given $X$ in the interactive data. 

\subsection{Decomposition of Treatment Variable}
After obtaining the representations of CIV ($Z_t$) and its conditional set ($Z_c$) from $X$, we use the CIV ($Z_t$) to reconstruct the treatment variable ($W$) and decompose $W$ to derive the causal relationship, specifically, user preference. We apply the least squares (LS) method based on IV4Rec~\cite{si2022model} to decompose $W$ and obtain the representation $\widehat{W}_{u,i}$ that is not affected by $U_c$:

\begin{equation}
    \widehat{W}_{u,i} = f_{pro}\left( Z_t,W_{u,i}\right),
\end{equation}
where $f_{pro}$ is the projection function that maps $Z_t$ and $W_{u,i}$ into the same space, allowing the unbiased  $\widehat{W}_{u,i}$. The function $f_{pro}(\cdot)$ is defined as: 
\begin{equation}
    f_{pro}\left( Z_t,W_{u,i} \right) = Z_t \cdot \mathcal{W}_{u,i},
\end{equation}
where $W_{u,i}$ is the closed-form solution of the LS method, and its calculation formula is as follows:
\begin{equation}
    \mathcal{W}_{u,i}= \arg \min_{W_{u,i}}\|Z_t \cdot W_{u,i} - W_{u,i}\|_2^2 = Z_t^\dagger \cdot W_{u,i},
\end{equation}
where $Z_t^\dagger$ is the Moore-Penrose pseudo-inverse of $Z_t$. Thus, the decomposed $W_{u, i}$ captures the causal relationship from $Z_t$ while separating the latent confounding bias introduced by $U_c$, which reflects user preferences. Additionally, we need to incorporate $Z_c$ to obtain the reconstructed $W_{u,i}^{re}$, as $Z_c$ blocks the confounding bias between $Z_t$ and $Y$. Therefore, our final reconstructed $W_{u,i}^{re}$ is obtained by:
\begin{equation}
    W_{u,i}^{re} =\alpha \widehat{W}_{u,i} + (1-\alpha)Z_c,
\end{equation}
where $\alpha$ is a hyperparameter used to balance $\widehat{W}_{u,i}$ and $Z_c$.

\subsection{Click Prediction}
To improve click prediction, we optimize the reconstructed treatment variables with BPR loss.
\begin{equation}
\begin{split}
L_{click} = & -\sum_{\left(u,p,n\right) \in D}  \ln sigma \big( 
\left<w_u^{re}, w_p^{re} \right> \\& - \left<w_u^{re}, w_n\right> \big),
\end{split}
\end{equation}
where $\left<\cdot\right>$ denotes the inner product, and $w_n$ is the negative sample item that user $u$ has not interacted with, selected by the Popularity-based Negative Sampling with Margin (PNSM) strategy~\cite{zheng2021disentangling}. By combining the ELBO and BPR, the final loss function of our IDCIV-RS is:
\begin{equation}
    L_{total} = L_{civ} + L_{click}.
\end{equation}

Therefore, our IDCIV-RS obtain $ W_{u,i}^{re}$ for click prediction by learning the representations of CIV $Z_t$ and its condition set $Z_c$ from $X$ and by decomposing $W$.

%Through the above steps, CIV4Rec obtain $\widehat{W}_{u, i}$  learns the representations of CIV $Z_t$, and its condition set $Z_c$ from $X$. It uses CIV decomposition to treatment variables to obtain unbiased causal relationship representation. At the same time, $\widehat{W}_{u, i}$ is fused with $Z_t$, which $Z_t$ contains latent confounding factor information to achieve click prediction.

\section{Experiments}
In this section, we conduct experiments on two real-world datasets to validate the recommendation performance and debiasing effectiveness of IDCIV-RS.

\subsection{Experimental Settings}
% We introduce the datasets, baselines, and metrics used in the experiments. 
% Detailed parameter settings and dataset preprocessing are provided in the Appendix due to the page limit.

\noindent\paragraph{Datasets.} We use two publicly real-world datasets: the Movielens-10M and the Douban-Movie datasets. Both datasets include user IDs, movie IDs, and ratings (1-5) for movies and are widely used in recommender system debiasing research. We use a dataset setup commonly used in the field of debiased recommender systems; specifically, the training set is biased and the test set is unbiased. We adopt the data pre-processing approach used in previous studies~\cite{zheng2021disentangling}.
% Additionally, we provide experimental results for the biased test set in the Appendix.

% We binarized the datasets based on movie ratings, assigning a value of 1 to five-star ratings and 0 to the rest, and applied ten-core filtering to the data.

% We follow the preprocessing steps used in DICE~\cite{zheng2021disentangling} for the dataset. First, we randomly select 40\% of the movie review data with the same probability (these data can be considered as \textit{unbiased} and represent recommendation results under a completely random strategy). Then, we use 10\% and 20\% of the random data as the validation set and test set, respectively, to evaluate the model's debiasing performance on unbiased data. Finally, the remaining 10\% of the \textit{random data} and the other 60\% of unselected samples form the training set. In other words,  the training, validation, and test data are split in a $7:1:2$ ratio. We include 10\% unbiased data in the training set due to the requirements of CausE~\cite{bonner2018causal}. Table~\ref{Table 1} shows the statistics of the processed datasets.

\begin{table}[h]     
    \centering
    \begin{tabular}{cccc} \hline
        Dataset & \# User & \# Item & \# Interaction  \\ \hline
        Movielens-10M & 37,962 & 4,819 & 1,371,473 \\ 
        Douban-Movie & 6,809 & 1,5012 & 173,766 \\ \hline
    \end{tabular}
    \caption{\centering Statistics of datasets.}
    \label{Table 1}
\end{table}

\begin{table*}[t]
		\centering
            \begin{tabular}{cc|cccccccc}
			\hline
			\multicolumn{2}{c|}{\multirow{2}{*}{Dataset}} &
			\multicolumn{8}{c}{Movielens-10M} \\ \cline{3-10}
			\multicolumn{2}{c|}{} &
			\multicolumn{4}{c|}{TopK=20} &
			\multicolumn{4}{c}{TopK=50} \\ \hline
			\multicolumn{1}{c|}{Backbone} &
			Method &
			\multicolumn{1}{c}{Recall$\uparrow$} &
			\multicolumn{1}{c}{HR$\uparrow$} &
			\multicolumn{1}{c}{NDCG$\uparrow$} &
			\multicolumn{1}{c|}{Imp.$\uparrow$} &
			\multicolumn{1}{c}{Recall$\uparrow$} &
			\multicolumn{1}{c}{HR$\uparrow$} &
			\multicolumn{1}{c}{NDCG$\uparrow$} &
			Imp.$\uparrow$ \\ \hline
			\multicolumn{1}{c|}{\multirow{9}{*}{MF}} &
			Original &
			\multicolumn{1}{c}{0.1276} &
			\multicolumn{1}{c}{0.4397} &
			\multicolumn{1}{c}{0.0832} &
			\multicolumn{1}{c|}{--} &
			\multicolumn{1}{c}{0.2332} &
			\multicolumn{1}{c}{0.6308} &
			\multicolumn{1}{c}{0.1156} &
			-- \\ %\cline{2-10} 
			\multicolumn{1}{c|}{} &
			IPS &
			\multicolumn{1}{c}{0.1228} &
			\multicolumn{1}{c}{0.4210} &
			\multicolumn{1}{c}{0.0779} &
			\multicolumn{1}{c|}{-3.76\%} &
			\multicolumn{1}{c}{0.2168} &
			\multicolumn{1}{c}{0.6016} &
			\multicolumn{1}{c}{0.1070} &
			-7.03\%\\ %\cline{2-10} 
			\multicolumn{1}{c|}{} &
			IPS-C &
			\multicolumn{1}{c}{0.1277} &
			\multicolumn{1}{c}{0.4335} &
			\multicolumn{1}{c}{0.0809} &
			\multicolumn{1}{c|}{+0.08\%} &
			\multicolumn{1}{c}{0.2224} &
			\multicolumn{1}{c}{0.6150} &
			\multicolumn{1}{c}{0.1102} &
			-4.63\%\\ %\cline{2-10} 
			\multicolumn{1}{c|}{} &
			CausE &
			\multicolumn{1}{c}{0.1164} &
			\multicolumn{1}{c}{0.4144} &
			\multicolumn{1}{c}{0.0770} &
			\multicolumn{1}{c|}{-8.77\%} &
			\multicolumn{1}{c}{0.2076} &
			\multicolumn{1}{c}{0.5940} &
			\multicolumn{1}{c}{0.1047} &
			-10.98\% \\ %\cline{2-10} 
			\multicolumn{1}{c|}{} &
			DICE &
			\multicolumn{1}{c}{{0.1626}} &
			\multicolumn{1}{c}{{0.5202}} &
			\multicolumn{1}{c}{{0.1076}} &
			\multicolumn{1}{c|}{{+27.42\%}} &
			\multicolumn{1}{c}{{0.2854}} &
			\multicolumn{1}{c}{{0.6941}} &
			\multicolumn{1}{c}{{0.1459}} &
			{+22.38\%} \\ %\cline{2-10} 
            \multicolumn{1}{c|}{} &
			DCCL &
			\multicolumn{1}{c}{0.1503} &
			\multicolumn{1}{c}{0.4874} &
			\multicolumn{1}{c}{0.0975} &
			\multicolumn{1}{c|}{+17.79\%} &
			\multicolumn{1}{c}{0.2636} &
			\multicolumn{1}{c}{0.6676} &
			\multicolumn{1}{c}{0.1326} &
			+13.04\% \\ %\cline{2-10}
            \multicolumn{1}{c|}{} &
			 IDCIV-RS-Causal &
			\multicolumn{1}{c}{\underline{0.1660}} &
			\multicolumn{1}{c}{\underline{0.5282}} &
			\multicolumn{1}{c}{\underline{0.1108}} &
			\multicolumn{1}{c|}{\underline{+30.09\%}} &
			\multicolumn{1}{c}{\underline{0.2895}} &
			\multicolumn{1}{c}{\underline{0.7012}} &
			\multicolumn{1}{c}{\underline{0.1495}} &
			\underline{+24.14\%} \\ 
		  \multicolumn{1}{c|}{} &
			 IDCIV-RS &
			\multicolumn{1}{c}{\pmb{0.1709}} &
			\multicolumn{1}{c}{\pmb{0.5362}} &
			\multicolumn{1}{c}{\pmb{0.1148}} &
			\multicolumn{1}{c|}{\pmb{+33.93\%}} &
			\multicolumn{1}{c}{\pmb{0.2973}} &
			\multicolumn{1}{c}{\pmb{0.7073}} &
			\multicolumn{1}{c}{\pmb{0.1542}} &
			\pmb{+27.49\%} \\ \hline
			\multicolumn{1}{c|}{\multirow{9}{*}{LightGCN}} &
			Original &
			\multicolumn{1}{c}{0.1462} &
			\multicolumn{1}{c}{0.4831} &
			\multicolumn{1}{c}{0.0952} &
			\multicolumn{1}{c|}{--} &
			\multicolumn{1}{c}{0.2631} &
			\multicolumn{1}{c}{0.6688} &
			\multicolumn{1}{c}{0.1316} &
			-- \\ %\cline{2-10} 
			\multicolumn{1}{c|}{} &
			IPS &
			\multicolumn{1}{c}{0.1298} &
			\multicolumn{1}{c}{0.4438} &
			\multicolumn{1}{c}{0.0849} &
			\multicolumn{1}{c|}{-11.22\%} &
			\multicolumn{1}{c}{0.2325} &
			\multicolumn{1}{c}{0.6196} &
			\multicolumn{1}{c}{0.1170} &
			-11.63\% \\ %\cline{2-10} 
			\multicolumn{1}{c|}{} &
			IPS-C &
			\multicolumn{1}{c}{0.1327} &
			\multicolumn{1}{c}{0.4533} &
			\multicolumn{1}{c}{0.0871} &
			\multicolumn{1}{c|}{-9.23\%} &
			\multicolumn{1}{c}{0.2383} &
			\multicolumn{1}{c}{0.6302} &
			\multicolumn{1}{c}{0.1201} &
			-9.43\% \\ %\cline{2-10}  
			\multicolumn{1}{c|}{} &
			CausE &
			\multicolumn{1}{c}{0.1164} &
			\multicolumn{1}{c}{0.4099} &
			\multicolumn{1}{c}{0.0727} &
			\multicolumn{1}{c|}{-20.38\%} &
			\multicolumn{1}{c}{0.2204} &
			\multicolumn{1}{c}{0.6080} &
			\multicolumn{1}{c}{0.1046} &
			-16.23\% \\ %\cline{2-10} 
			\multicolumn{1}{c|}{} &
			DICE &
			\multicolumn{1}{c}{\underline{0.1810}} &
			\multicolumn{1}{c}{\underline{0.5564}} &
			\multicolumn{1}{c}{\underline{0.1228}} &
			\multicolumn{1}{c|}{\underline{+23.80\%}} &
			\multicolumn{1}{c}{\underline{0.3109}} &
			\multicolumn{1}{c}{\underline{0.7219}} &
			\multicolumn{1}{c}{\underline{0.1632}} &
			\underline{+18.17\%}\\ %\cline{2-10} 
            \multicolumn{1}{c|}{} &
			DCCL &
			\multicolumn{1}{c}{0.1462} &
			\multicolumn{1}{c}{0.4824} &
			\multicolumn{1}{c}{0.0947} &
			\multicolumn{1}{c|}{{0\%}} &
			\multicolumn{1}{c}{0.2644} &
			\multicolumn{1}{c}{0.6711} &
			\multicolumn{1}{c}{0.1311} &
			+0.49\%\\ %\cline{2-10}
            \multicolumn{1}{c|}{} &
			IDCIV-RS-Causal &
			\multicolumn{1}{c}{{0.1784}} &
			\multicolumn{1}{c}{{0.5511}} &
			\multicolumn{1}{c}{{0.1205}}&
			\multicolumn{1}{c|}{{+22.02\%}} &
			\multicolumn{1}{c}{{0.3056}} &
			\multicolumn{1}{c}{{0.7160}} &
			\multicolumn{1}{c}{{0.1602}} &
			{+16.15\%}\\ %\cline{2-10}
			\multicolumn{1}{c|}{} &
			 IDCIV-RS &
			\multicolumn{1}{c}{\pmb{0.1817}} &
			\multicolumn{1}{c}{\pmb{0.5582}} &
			\multicolumn{1}{c}{\pmb{0.1241}} &
			\multicolumn{1}{c|}{\pmb{+24.28\%}} &
			\multicolumn{1}{c}{\pmb{0.3119}} &
			\multicolumn{1}{c}{\pmb{0.7232}} &
			\multicolumn{1}{c}{\pmb{0.1645}} &
			\pmb{+18.55\%}\\ \hline
		\end{tabular}
        \caption{The performance of all methods on Movielens-10M. The ``original" indicates that only the backbone is used, with no additional causal debiasing methods. The best results are highlighted in bold, and the second-best results are underlined. }
    
		\label{Table 2}	
     \end{table*}

     \begin{table*}[!h]
		\centering	
		\begin{tabular}{cc|cccccccc}
			\hline
			\multicolumn{2}{c|}{\multirow{2}{*}{Dataset}} &
			\multicolumn{8}{c}{Douban-Movie} \\ \cline{3-10} 
			\multicolumn{2}{c|}{} &
			\multicolumn{4}{c|}{TopK=20} &
			\multicolumn{4}{c}{TopK=50} \\ \hline
			\multicolumn{1}{c|}{Backbone} &
			Method &
			\multicolumn{1}{c}{Recall$\uparrow$} &
			\multicolumn{1}{c}{HR$\uparrow$} &
			\multicolumn{1}{c}{NDCG$\uparrow$} &
			\multicolumn{1}{c|}{Imp.$\uparrow$} &
			\multicolumn{1}{c}{Recall$\uparrow$} &
			\multicolumn{1}{c}{HR$\uparrow$} &
			\multicolumn{1}{c}{NDCG$\uparrow$} &
			Imp.$\uparrow$ \\ \hline
			\multicolumn{1}{c|}{\multirow{9}{*}{MF}} &
			Original &
			\multicolumn{1}{c}{0.0214} &
			\multicolumn{1}{c}{0.0542} &
			\multicolumn{1}{c}{0.0128} &
			\multicolumn{1}{c|}{--} &
			\multicolumn{1}{c}{0.0371} &
			\multicolumn{1}{c}{0.0933} &
			\multicolumn{1}{c}{0.0171} &
			-- \\ %\cline{2-10} 
			\multicolumn{1}{c|}{} &
			IPS &
			\multicolumn{1}{c}{0.0172} &
			\multicolumn{1}{c}{0.0444} &
			\multicolumn{1}{c}{0.0099} &
			\multicolumn{1}{c|}{-19.63\%} &
			\multicolumn{1}{c}{0.0282} &
			\multicolumn{1}{c}{0.0755} &
			\multicolumn{1}{c}{0.0130} &
			-23.99\%\\ %\cline{2-10} 
			\multicolumn{1}{c|}{} &
			IPS-C &
			\multicolumn{1}{c}{0.0166} &
			\multicolumn{1}{c}{0.0446} &
			\multicolumn{1}{c}{0.0095} &
			\multicolumn{1}{c|}{-22.43\%} &
			\multicolumn{1}{c}{0.0271} &
			\multicolumn{1}{c}{0.0761} &
			\multicolumn{1}{c}{0.0125} &
			-26.95\%\\ %\cline{2-10}  
			\multicolumn{1}{c|}{} &
			CausE &
			\multicolumn{1}{c}{0.0149} &
			\multicolumn{1}{c}{0.0410} &
			\multicolumn{1}{c}{0.0074} &
			\multicolumn{1}{c|}{-30.37\%} &
			\multicolumn{1}{c}{0.0273} &
			\multicolumn{1}{c}{0.0761} &
			\multicolumn{1}{c}{0.0108} &
			-26.42\%\\ %\cline{2-10} 
			\multicolumn{1}{c|}{} &
			DICE &
			\multicolumn{1}{c}{{0.0231}} &
			\multicolumn{1}{c}{{0.0615}} &
			\multicolumn{1}{c}{{0.0133}} &
			\multicolumn{1}{c|}{{+7.94\%}} &
			\multicolumn{1}{c}{{0.0396}} &
			\multicolumn{1}{c}{{0.1012}} &
			\multicolumn{1}{c}{{0.0178}} &
			{+6.74\%} \\ %\cline{2-10}
            \multicolumn{1}{c|}{} &
			DCCL &
			\multicolumn{1}{c}{0.0217} &
			\multicolumn{1}{c}{0.0595} &
			\multicolumn{1}{c}{0.0123} &
			\multicolumn{1}{c|}{+1.40\%} &
			\multicolumn{1}{c}{0.0385} &
			\multicolumn{1}{c}{0.1040} &
			\multicolumn{1}{c}{0.0170} &
			+3.77\%\\ %\cline{2-10}
            \multicolumn{1}{c|}{} &
			 IDCIV-RS-Causal &
			\multicolumn{1}{c}{\underline{0.0278}} &
			\multicolumn{1}{c}{\underline{0.0736}} &
			\multicolumn{1}{c}{\underline{0.0162}} &
			\multicolumn{1}{c|}{\underline{+29.91\%}} &
			\multicolumn{1}{c}{\underline{0.0462}} &
			\multicolumn{1}{c}{\underline{0.1213}} &
			\multicolumn{1}{c}{\underline{0.0213}} &
			\underline{+24.53\%} \\ 
			\multicolumn{1}{c|}{} &
			 IDCIV-RS &
			\multicolumn{1}{c}{\pmb{0.0299}} &
			\multicolumn{1}{c}{\pmb{0.0777}} &
			\multicolumn{1}{c}{\pmb{0.0171}} &
			\multicolumn{1}{c|}{\pmb{+39.71\%}} &
			\multicolumn{1}{c}{\pmb{0.0480}} &
			\multicolumn{1}{c}{\pmb{0.1213}} &
			\multicolumn{1}{c}{\pmb{0.0220}} &
			\pmb{+29.38\%} \\ \hline
			\multicolumn{1}{c|}{\multirow{9}{*}{LightGCN}} &
			Original &
			\multicolumn{1}{c}{0.0375} &
			\multicolumn{1}{c}{0.0557} &
			\multicolumn{1}{c}{0.0118} &
			\multicolumn{1}{c|}{--} &
			\multicolumn{1}{c}{0.0640} &
			\multicolumn{1}{c}{0.0908} &
			\multicolumn{1}{c}{0.0155} &
			--\\ %\cline{2-10} 
			\multicolumn{1}{c|}{} &
			IPS &
			\multicolumn{1}{c}{0.0352} &
			\multicolumn{1}{c}{0.0928} &
			\multicolumn{1}{c}{0.0208} &
			\multicolumn{1}{c|}{-6.13\%} &
			\multicolumn{1}{c}{0.0619} &
			\multicolumn{1}{c}{0.1576} &
			\multicolumn{1}{c}{0.0281} &
			-3.28\%\\ %\cline{2-10} 
			\multicolumn{1}{c|}{} &
			IPS-C &
			\multicolumn{1}{c}{0.0368} &
			\multicolumn{1}{c}{0.0980} &
			\multicolumn{1}{c}{0.0219} &
			\multicolumn{1}{c|}{-1.87\%} &
			\multicolumn{1}{c}{0.0643} &
			\multicolumn{1}{c}{0.1657} &
			\multicolumn{1}{c}{0.0295} &
			-4.69\%\\ %\cline{2-10}  
			\multicolumn{1}{c|}{} &
			CausE &
			\multicolumn{1}{c}{0.0263} &
			\multicolumn{1}{c}{0.0693} &
			\multicolumn{1}{c}{0.0143} &
			\multicolumn{1}{c|}{-29.87\%} &
			\multicolumn{1}{c}{0.0463} &
			\multicolumn{1}{c}{0.1216} &
			\multicolumn{1}{c}{0.0199} &
			-27.66\%\\ %\cline{2-10} 
			\multicolumn{1}{c|}{} &
			DICE &
			\multicolumn{1}{c}{{0.0401}} &
			\multicolumn{1}{c}{{0.1088}} &
			\multicolumn{1}{c}{{0.0232}} &
			\multicolumn{1}{c|}{{+6.93\%}} &
			\multicolumn{1}{c}{{0.0679}} &
			\multicolumn{1}{c}{{0.1755}} &
			\multicolumn{1}{c}{{0.0310}} &
			{+6.09\%} \\ %\cline{2-10} 
            \multicolumn{1}{c|}{} &
			DCCL &
			\multicolumn{1}{c}{{0.0401}} &
			\multicolumn{1}{c}{0.1046} &
			\multicolumn{1}{c}{0.0225} &
			\multicolumn{1}{c|}{+6.93\%} &
			\multicolumn{1}{c}{0.0693} &
			\multicolumn{1}{c}{0.1732} &
			\multicolumn{1}{c}{0.0306} &
			+8.28\%\\ %\cline{2-10}
			\multicolumn{1}{c|}{} &
			IDCIV-RS-Causal &
			\multicolumn{1}{c}{\underline{0.0407}} &
			\multicolumn{1}{c}{\underline{0.1125}} &
			\multicolumn{1}{c}{\underline{0.0239}} &
			\multicolumn{1}{c|}{\underline{+8.53\%}} &
			\multicolumn{1}{c}{\underline{0.0704}} &
			\multicolumn{1}{c}{\underline{0.1840}} &
			\multicolumn{1}{c}{\underline{0.0321}} &
			\underline{+10.00\%}\\ 
            \multicolumn{1}{c|}{} &
                IDCIV-RS &
			\multicolumn{1}{c}{\pmb{0.0435}} &
			\multicolumn{1}{c}{\pmb{0.1178}} &
			\multicolumn{1}{c}{\pmb{0.0252}} &
			\multicolumn{1}{c|}{\pmb{+16.00\%}} &
			\multicolumn{1}{c}{\pmb{0.0724}} &
			\multicolumn{1}{c}{\pmb{0.1886}} &
			\multicolumn{1}{c}{\pmb{0.0332}} &
			\pmb{+13.13\%}\\ \hline
		\end{tabular}
        \caption{The results of all methods on Douban-Movie. The best is highlighted in bold, and the second-best is underlined.
        }
		\label{Table 3}
\end{table*}
     
\noindent\paragraph{Baselines.}
% Causal debiasing methods are typically used as adjunct techniques alongside a backbone recommendation model. In our experiments, we use the mainstream recommendation models, MF and LightGCN, as the backbone. We compare our approach with the following five debiased recommendation methods based on causality:
Causal debiasing methods are typically applied as enhancements to backbone recommendation models. In our experiments, we use MF and LightGCN as the backbone models. We compare our approach against five causality-based debiasing methods:
\begin{itemize}
    \item[\textbullet] \textbf{IPS}~\cite{schnabel2016recommendations}: This method assigns weights that are the inverse of an item’s popularity, thereby enhancing the impact of less popular items while reducing the influence of more popular ones.
    \item[\textbullet] \textbf{IPS-C}~\cite{bottou2013counterfactual}: This approach caps the maximum value of IPS weights to reduce variance across the entire weight distribution.
    \item[\textbullet] \textbf{CausE}~\cite{bonner2018causal}: This method generates two sets of embeddings from the data, which are then aligned using regularization techniques to ensure their similarity.
    \item[\textbullet] \textbf{DICE}~\cite{zheng2021disentangling}: This method uses Structural Causal Modeling (SCM)~\cite{pearl2009causality} to define user-item interactions. This approach leverages the collision effect of causal reasoning to enhance training effectiveness.
    \item[\textbullet] \textbf{DCCL}~\cite{zhao2023disentangled}: This method uses contrastive learning to address data sparsity and the separation of these components.
\end{itemize}

We did not compare IDCIV-RS with IV-based methods like IV4Rec~\cite{si2022model}, as these require explicit IVs derived from domain knowledge or user search data, which our datasets lack. In contrast, IDCIV-RS learns CIV representations directly from user interactions, avoiding dependence on unavailable or domain-specific data. This confers greater flexibility and applicability in real-world scenarios.
% We did not compare our IDCIV-RS method with IV-based debiasing methods like IV4Rec~\cite{si2022model}, as these methods require specifying an IV based on domain knowledge or expert input, such as self-collected user search data, which is not available in the experimental datasets. In contrast, IDCIV-RS learns the representations of the CIV and its conditional set directly from user interaction data. The datasets used in our experiments lack such explicit IVs, making a meaningful comparison unfeasible. This highlights the advantage of IDCIV-RS, which does not rely on potentially unavailable or domain-specific data, offering greater flexibility and applicability in real-world scenarios.

\paragraph{Metrics.} We evaluate Top-K recommendation under implicit feedback using Recall, Hit Rate (HR), and NDCG. Results reflect each method’s best performance under optimal settings. ``Imp." denotes the percentage improvement in Recall over the base model.

\subsection{Comparison of Experimental Results}
Tables \ref{Table 2} and \ref{Table 3} present the results of IDCIV-RS and all baseline approaches on two real-world datasets. \textit{IDCIV-RS-Causal} is a variant of IDCIV-RS that denotes click prediction using only the unbiased embeddings $\widehat{W}_{u, i}$, which capture the causal relationship.
	 
The analysis of Tables~\ref{Table 2} and~\ref{Table 3} shows that IDCIV-RS and IDCIV-RS-Causal significantly improve performance metrics compared to the original backbone, with the highest improvement reaching 39.71\%, demonstrating statistical significance and the superiority of our approach. Notably, IDCIV-RS consistently outperforms IDCIV-RS-Causal, aligning with the understanding that incorporating appropriate confounders enhances recommendation performance. This confirms that $Z_c$ in IDCIV-RS effectively captures relevant confounders in user interaction data.

Tables~\ref{Table 2} and~\ref{Table 3} provide several key insights: (1) IPS-based debiasing methods perform poorly due to their reliance on the inverse propensity score, which is sensitive to data distribution. In our experiments, training on a biased dataset and testing on an unbiased one led to distribution mismatches. (2) CausE also underperforms, as it requires an unbiased training dataset to align user-item embeddings. (3) Although DICE and DCCL, which are based on causal graph assumptions, improve performance, they still fall short of optimal results. This is because they target specific biases based on predefined causal graphs, while real-world datasets often contain diverse biases from latent confounders, limiting their effectiveness.

% Tables~\ref{Table 2} and~\ref{Table 3} reveal several key insights: (1) IPS-based debiasing methods perform poorly because they rely heavily on the inverse propensity score, which is closely tied to the data distribution. In our experiments, training was conducted on a biased dataset, while testing was performed on an unbiased dataset, leading to inconsistent data distributions between the training and validation sets. (2) Similarly, CausE showed suboptimal performance because it requires an unbiased dataset during training to align the user-item embedding representations. (3) Although DICE and DCCL, two debiasing methods based on causal graph assumptions, improve recommendation performance, they do not achieve optimal results. This is because these methods target specific biases based on assumed causal graphs, while real-world datasets often contain diverse biases due to latent confounding factors, limiting their effectiveness.

% Note that on the Movielens-10M dataset, while CIV4Rec's improvement with LightGCN appears marginal compared to DICE, it is statistically significant, with a $p$-value of less than $0.05$.

% On the Movielens-10M dataset, IDCIV-RS with LightGCN shows a marginal but statistically significant improvement over DICE ($p$-value $<$ 0.05). This can be attributed to IDCIV-RS’s ability to learn the CIV and its conditional set from interactive data, which reduces the typical constraints of such methods and effectively mitigates bias, leading to improved recommendation performance.

\subsection{Evaluation on Debiasing Experiments Ability}
% In this section, we evaluate the debiasing ability of CIV4Rec. We calculate the proportion of popular items recommended by all methods and use Intersection Over Union (IOU)~\cite{zheng2021disentangling} as the evaluation indicator. A higher IOU value means a higher proportion of popular items in the recommendations, indicating weaker debiasing ability.

We use the \textit{Intersection Over Union} (\textit{IOU})~\cite{zheng2021disentangling} metric to evaluate the debiasing ability of all methods. A higher IOU reflects more popular items in the recommendations, indicating weaker debiasing performance.

\begin{figure}[h]
    \centering
    \includegraphics[width=\columnwidth]{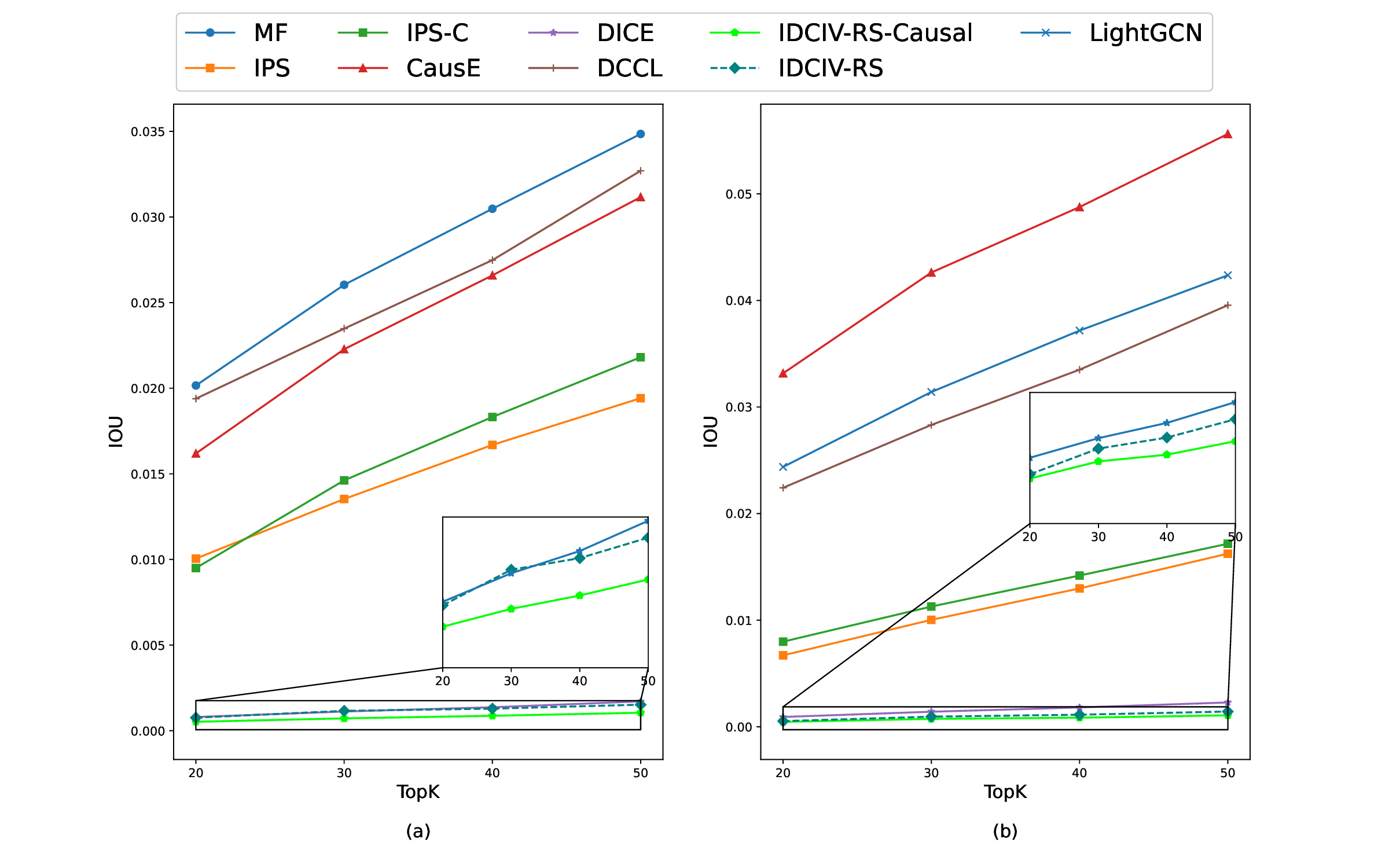}
    \caption{The IOU of recommended items and popular items for all methods on the Douban-movie dataset. (a) IOU of all methods on the MF; (b) IOU of all methods on the LightGCN.}
    \label{Douban_IOU}
\end{figure}

Figure~\ref{Douban_IOU} shows the IOU for all methods on the Douban-movie dataset. IDCIV-RS and IDCIV-RS-Causal exhibit the lowest IOU, indicating superior debiasing ability. Notably, the IOU for all baseline methods increases significantly as the number of recommended items grows, suggesting that their debiasing ability diminishes with more recommendations. In contrast, the debiasing ability of our IDCIV-RS and IDCIV-RS-Causal remains relatively stable, demonstrating greater robustness.

Figure~\ref{Douban_IOU} also illustrates that the IOU of IDCIV-RS is higher than that of IDCIV-RS-Causal, due to IDCIV-RS incorporating confounding factor information. This suggests that $Z_c$ effectively captures confounding factors in user interaction data, validating our method. 

\subsection{Ablation Studies}
\begin{figure}[h]
    \centering
    \includegraphics[width=\columnwidth]{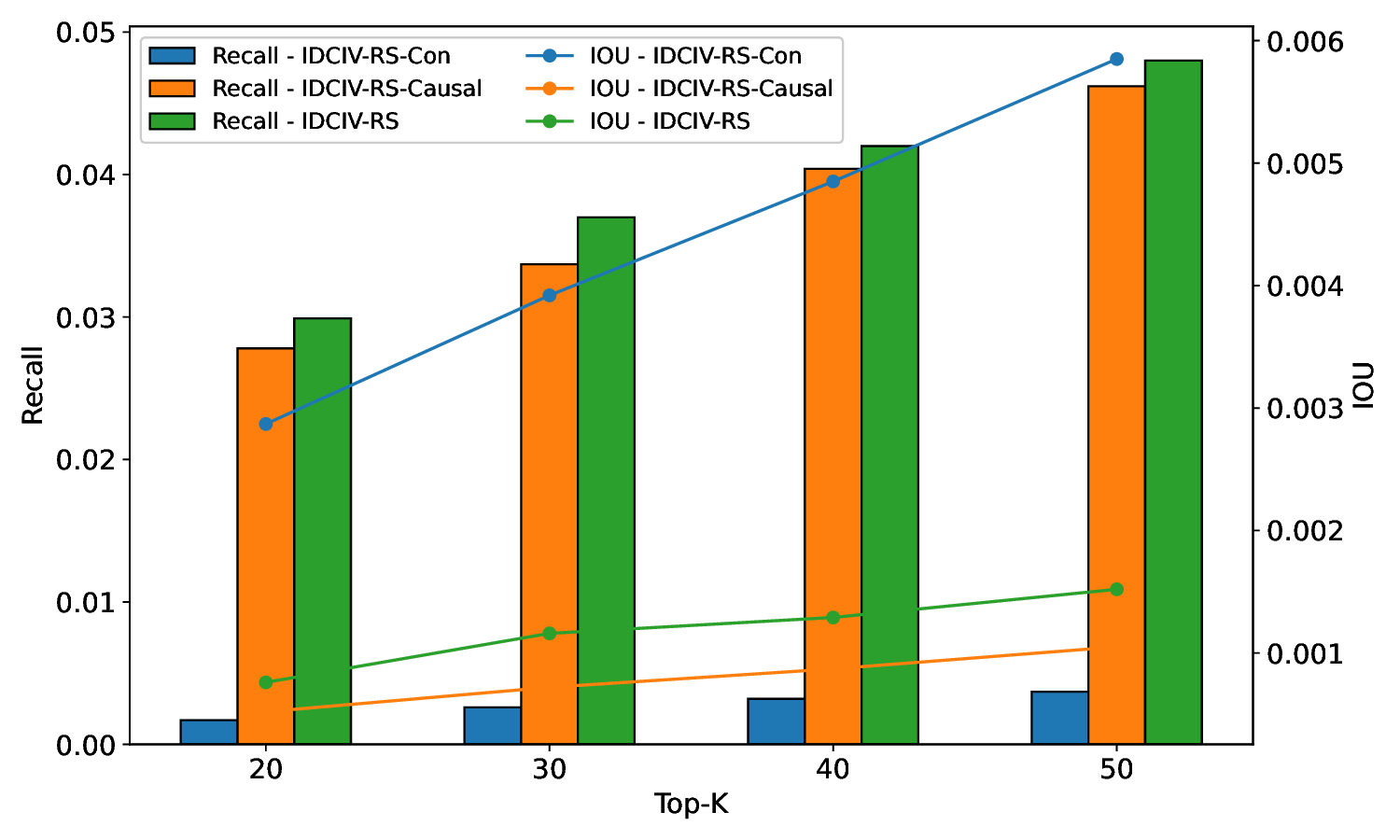}
    \caption{Recall and IOU of MF-based IDCIV-RS and its variants on the Douban-movie dataset. Where the bar represents Recall and the curve represents IOU.}
    \label{Ablation}
\end{figure}

We perform ablation studies to evaluate the effectiveness of each component in IDCIV-RS. To verify the effectiveness of CIV and its condition set, we propose \textit{IDCIV-RS-Con}, which uses only $Z_c$ for click prediction. Figure~\ref{Ablation} presents the IOU and Recall of IDCIV-RS and its variants. The results show that IDCIV-RS-Con has the highest IOU and lowest Recall, highlighting the effectiveness of $Z_c$ in capturing confounding factor information. In contrast, IDCIV-RS-Causal exhibits higher Recall but lower IOU than IDCIV-RS-Con, indicating its effectiveness in capturing user preference information and mitigating confounding factors through $Z_t$. IDCIV-RS, by integrating both user preference and confounding factor information, achieves higher Recall and IOU than IDCIV-RS-Causal, demonstrating the combined effectiveness of $Z_t$ and $Z_c$.

% To verify the effectiveness of CIV and its condition set learned by CIV4Rec, we propose a variant: CIV4Rec-Con, which only uses $Z_c$ for click prediction. Figure~\ref{Ablation} shows the IOU and Recall of CIV4Rec and its variants. From the figure, we see that CIV4Rec-Con has the highest IOU and the lowest Recall. This shows that CIV4Rec-Con captures the confounding factor information in user interaction data, strongly verifying the effectiveness of $Z_c$.

% In contrast, the Recall of CIV4Rec-Causal is significantly higher than that of CIV4Rec-Con, while the IOU is significantly lower. This indicates that CIV4Rec-Causal captures user preference information and reduces the impact of confounding factors, verifying the effectiveness of CIV $Z_t$.

% CIV4Rec integrates both user preference information and confounding factor information, so its Recall and IOU are both higher than those of CIV4Rec-Causal. This fully illustrates the effectiveness of the CIV $Z_t$ and its condition set $Z_c$ learned by our method.

\section{Conclusion}
In this paper, we propose a data-driven CIV debiasing method called IDCIV-RS. We learn the representations of CIV and its conditional set from user interaction data. The CIV is used to decompose the treatment variable and uncover the causal relationships between variables, while the conditional set captures confounding factors in the user interaction data. Unlike existing IV-based debiasing methods, IDCIV-RS imposes fewer constraints and does not require the selection of specific IVs based on domain knowledge. By integrating confounding factors and the causal relationships of the treatment variable, IDCIV-RS achieves high-quality recommendations and effective debiasing. We conducted extensive experiments on two real-world datasets to validate the effectiveness and superiority of IDCIV-RS in both recommendation and debiasing performance.

\section*{Acknowledgments}
This work is supported partly by the Project of Guangxi Science and Technology (2025GXNSFFA069015), the National Natural Science Foundation of China (No. 62372119, 62166003), the Key Lab of Education Blockchain and Intelligent Technology, Ministry of Education (No. EBME24-03), the Research Fund of the Guangxi Key Lab of Multi-source Information Mining \& Security (MIMS24-M-01), and the Australian Research Council (under grant DP230101122).

\bibliographystyle{named}
\bibliography{ijcai25}

\end{document}